\documentclass[aps,prb,preprint,groupedaddress]{revtex4}

\usepackage[pdftex]{graphicx}
\begin{document}


\title{Atomic and electronic structure of monolayer graphene on $6H$-SiC$(000\overline{1}) (3\times3)$ : a scanning tunneling microscopy study.}


\author{F. Hiebel, P. Mallet, L. Magaud and J.-Y. Veuillen}
\affiliation{Institut N\'{e}el, CNRS-UJF, Bo\^{i}te Postale 166, 38042 Grenoble Cedex 9, France}


\date{\today}

\begin{abstract}
We present an investigation of the atomic and electronic structure of graphene monolayer islands on the $6H$-SiC$(000\overline{1})(3\times3)$ (SiC$(3\times3)$) surface reconstruction using scanning tunneling microscopy (STM) and spectroscopy (STS).
The orientation of the graphene lattice changes from one island to the other. In the STM images, this rotational disorder gives rise to various superlattices with periods in the nm range. We show that those superlattices are moir\'{e} patterns (MPs) and we correlate their apparent height with the stacking at the graphene/SiC$(3\times3)$ interface. The contrast of the MP in STM images corresponds to a small topographic modulation (by typically 0.2 \AA) of the graphene layer. From STS measurements we find that the substrate surface presents a $1.5$ eV wide bandgap encompassing the Fermi level. This substrate surface bandgap subsists below the graphene plane. The tunneling spectra are spatially homogeneous on the islands within the substrate surface gap, which shows that the MPs do not impact the low energy electronic structure of graphene. We conclude that the SiC$(3\times3)$ reconstruction efficiently passivates the substrate surface and that the properties of the graphene layer which grows on top of it should be similar to those of the ideal material.
\end{abstract}


\maketitle

\section{Introduction}
Fascinating properties have been predicted and observed for monolayer graphene\cite{rise, neto:109}. Among them one finds the anomalous quantum Hall effect\cite{Nov05, Kim05}, the Klein tunneling phenomenon\cite{KleinKatsnelson, KleinKim}, and weak (anti)localization effects\cite{wu:136801, mccann:146805}. Moreover, suspended graphene shows exceptionally high carriers mobility\cite{bolotin, andrei} even near room temperature\cite{PKimPRL}. These features make graphene an attractive material for the investigation of original physical phenomena\cite{Veselago, Billiard} and for the development of devices such as transistors\cite{FieldEffectNov, FETKim} and captors\cite{detection}. 

The physical properties of free standing graphene are intimately linked to the presence of two equivalent carbon sublattices commonly called A and B. Usually, graphene layers are supported on a substrate and the interaction between the electronic states of the substrate surface and the $p_z$ orbitals of the C atoms can significantly alter the electronic structure -and thus the properties- of the material. This has been shown recently by angle resolved photoemission for graphene elaborated on metal surfaces where this coupling modifies the band dispersion close to the Dirac point\cite{GNi, brugger:045407, SutterNano}, suppressing the ``Dirac cones''. The investigation of the atomic and electronic structure of the interface between graphene and the substrate is thus of primary importance. This is in particular the case for few layers graphene grown on SiC substrates, where as-grown samples are used for physical measurements \cite{CB04, CB, wu:136801, revuedeHeer}, since the doped graphene layers close to the interface should give the largest contribution to electrical transport\cite{CB}.
 
Few layers graphene are obtained by high temperature treatment of the polar faces of SiC substrates \cite{vanBommel, PhysRevB.58.16396, interactionForbeaux}. Usually commercial hexagonal ($4H$ or $6H$) substrates are used. They have two different faces, the $(0001)$ one (the Si face) and the $(000\overline{1})$ faces (the C face). The interface between the Si face and the graphene overlayer has been extensively studied in the last few years. The current model for this interface is that the first graphitic layer strongly interacts with the substrate, giving rise to the $(6\sqrt{3}\times6\sqrt{3})R30^\circ$ ($6R3$) reconstruction\cite{kimPRL, varchon:235412}. Covalent bonds form between Si atoms of the substrate surface and the graphene layer, which results in the suppression of the Dirac cones characteristic of graphene \cite{mattausch:076802, varchon:126805, kimPRL}. This model is supported by photoemission data\cite{emtsev:155303}. Accordingly no graphene contrast  has been detected in STM images of the $6R3$ reconstruction\cite{mallet:041403, riedl:245406, rutter:235416, lauffer}, which is usually called the ``buffer layer''. The electronic structure of graphene is developed only for the second C plane\cite{kimPRL, varchon:235412, varchon:126805, mattausch:076802}, where a band structure very similar to the Dirac cones has been observed experimentally\cite{NatureBostwick, Zhou}. The question of a possible perturbation of the electronic structure of the graphene layer due to an interaction with the buffer layer remains open\cite{gapRotenberg, gaplanzara}. Nevertheless, the honeycomb contrast expected for ideal graphene is observed by STM on this second C plane\cite{mallet:041403, rutter:235416, lauffer, brar:122102}. Moreover the analysis of the standing wave patterns indicates that the electronic chirality of graphene is preserved\cite{brihuega:206802}. 

The interface between graphene and the C face has been less extensively studied. It has long been known that the growth is quite different on the C and the Si face\cite{vanBommel}. Graphitic films grown in UHV conditions on the C face exhibit some rotational disorder\cite{vanBommel, interactionForbeaux}. This disorder already exists for the first C layer\cite{emtsev:155303, hiebel:153412, revstarke}. Interestingly, it was found using photoemission that the interaction between the first C layer and the substrate was much weaker than for the Si face: no buffer layer is detected in core level spectroscopy\cite{revstarke, emtsev:155303} and the band structure of this layer\cite{emtsev:155303} resembles the one of graphene. The situation is however complicated by the facts that i) two different pristine reconstructions of the substrate -the SiC$(2\times2)_C$ and the SiC$(3\times3)$- exist at the interface below the graphene layer\cite{revstarke, emtsev:155303, hiebel:153412} and ii) that several orientations exist for the graphene islands for each reconstruction, leading to different superlattices\cite{hiebel:153412}. A systematic analysis of the interface for the two different substrate reconstructions aiming at understanding their atomic and electronic structure for the different orientations of the graphene layer is thus needed. This is best achieved by STM, which can address each individual island.

In a previous paper we have shown that a graphitic signal could be observed at low bias for both the SiC$(2\times2)_C$ and the SiC$(3\times3)$ reconstructions, indicating a weaker coupling with the substrate than on the Si face\cite{hiebel:153412}. A recent ab-initio calculation has shown that the reduced interaction in the case of the SiC$(2\times2)_C$ reconstruction is due to a passivation of the substrate surface by Si adatoms\cite{magaud:161405}. The linear dispersion of the graphene bands close to the Dirac point is preserved, but a residual coupling with the substrate, also evidenced by STM, was found. In the present paper we concentrate on the graphene islands formed on the SiC$(3\times3)$ interface reconstruction, which are called $G\_3\times3$ islands afterward, where the interaction with the substrate seems to be even smaller\cite{hiebel:153412}. We first analyze the geometric structure of the superlattices. We show that they are moir\'{e} patterns and we relate their apparent height to the local stacking at the interface. We then analyze the electronic structure of the $G\_3\times3$ islands compared to that of the bare substrate SiC$(3\times3)$ reconstruction. A wide surface bandgap (of width $\approx1.5$ eV) is found by STS in the electronic structure of the bare reconstruction, which persists below the graphene layer. The Fermi level of graphene is located in the vicinity of the top of this surface gap. In STM images a graphene signal dominates inside the substrate surface gap, and the STS data explain the high bias ``transparency'' of graphene. A comparison between STM images of $G\_3\times3$ islands for a specific orientation with previous ab-initio calculations -as well as with the case of the Si face- indicates that the substrate reconstruction is responsible for the weak graphene-substrate interaction. Finally we show that the moir\'{e} pattern is essentially of topographic origin. It is associated with small undulations of the graphene layer. From STS, these undulations do not lead to heterogeneities in the electronic structure of graphene, at variance with the case of more strongly interacting systems such as graphene on Ru(0001)\cite{parga:056807}. From these data we conclude that the $G\_3\times3$ islands should be a system close to ideal, uncoupled, graphene. At present it is not clear whether the $G\_3\times3$ structure is present at the interface for few layers graphene films elaborated at high temperature in non UHV conditions\cite{ConradPRB}. Our results indicate anyway that manipulating the atomic structure of the surface can be a useful way to modify the coupling at the interface, as shown previously for metal substrates\cite{varykhalov:157601}. 

\section{Experiment}

The sample preparation and characterization were conducted under ultrahigh vacuum. The sample graphitization was performed in-situ by following the procedure presented previously\cite{hiebel:153412}. The n doped $6H$-SiC$(000\overline{1})$ samples were first cleaned by annealing under a Si flux at $850^\circ$C. The already reported SiC$(3\times3)$ reconstruction\cite{Hoster3x3} was obtained by further annealing at $950 -1000^\circ$C. After annealing steps at increasing temperature, a graphene coverage of less than a monolayer is finally detected by low energy electron diffraction (LEED) and Auger spectroscopy. At this stage, the LEED patterns show SiC$(3\times3)$ and SiC$(2\times2)$ spots and a ring-shaped graphitic signal with modulated intensity\cite{emtsev:155303, hiebel:153412, revstarke}. 

The STM and STS measurements were made at room temperature with mechanically cut PtIr tips. 5 samples were investigated, using more than 10 macroscopically different tips.
The samples morphology observed by STM was similar to previous results, with the presence of bare SiC$(3\times3)$ reconstructed substrate domains, graphene monolayer islands on the SiC$(3\times3)$ reconstruction (G\_$3\times3$) and on the SiC$(2\times2)_C$ reconstruction (G\_$2\times2$) and also few multilayer islands\cite{hiebel:153412, revstarke}. The focus of this paper is the structure of G\_$3\times3$ islands and several dozens of them were observed.

\section{Results and discussion}

\subsection{\label{geom}Superlattices and local stacking of monolayer graphene on the SiC$(3\times3)$ reconstruction}

\begin{figure*}[h!]
\includegraphics[width=0.8\textwidth]{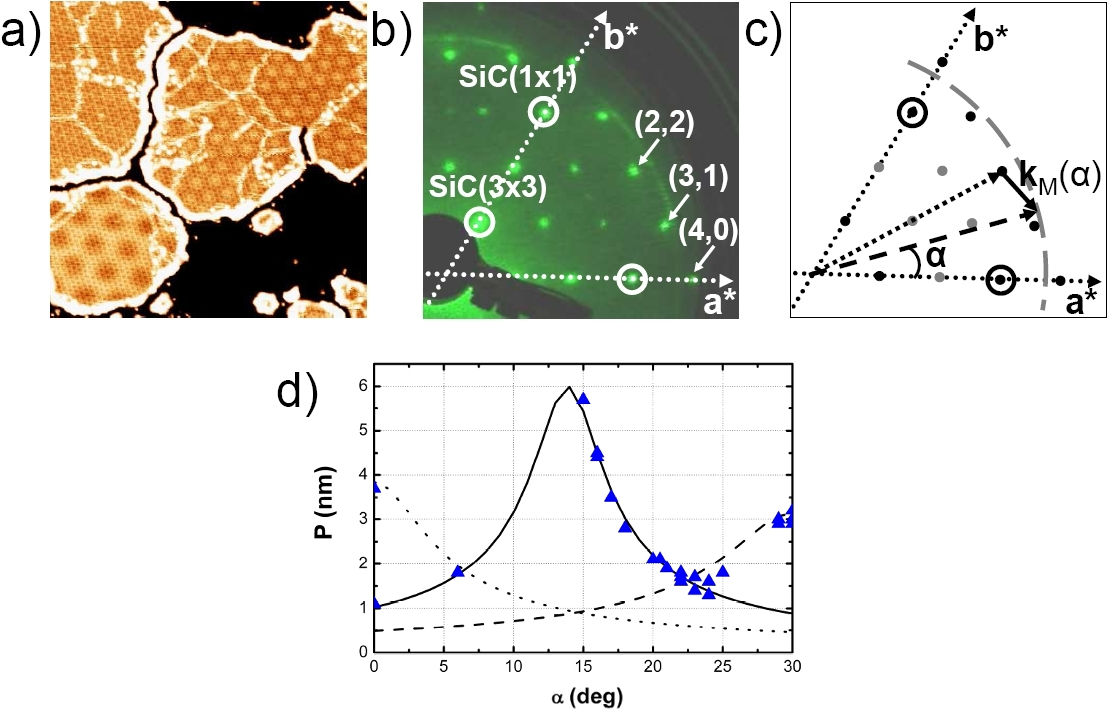}
 \caption{\label{fig1} (Color online) (a) $50\times50$ nm$^2$ STM image of $G\_(3\times3)$ islands on SiC$(000\overline{1})$ with enhanced contrast to reveal their superlattices. Sample bias: $-2.5$ V. (b) LEED pattern of the sample, primary energy: $78$ eV. Circles indicate first order SiC$(1\times1)$ and SiC$(3\times3)$ spots. The relevant high order SiC$(3\times3)$ spots close to the graphitic ring-shaped signal are indicated by arrows and their coordinates are given. (c) Schematic picture of the LEED pattern in (b). The dashed curve stands for the graphitic signal. The dashed arrow is the reciprocal lattice vector that corresponds to a graphene layer rotated by an angle $\alpha$ with respect to the SiC$(1\times1)$ surface. The dotted one corresponds to the $(2,2)$ reciprocal vector of the SiC$(3\times3)$ reconstruction. The reciprocal lattice vector of the moir\'{e} pattern $\mathbf{k_M}(\alpha)$ constructed on these vectors is represented by a solid vector. (d) The calculated moir\'{e} periods $P(\alpha)$ for (4,0), (3,1) and (2,2) SiC$(3\times3)$ reciprocal lattice vectors are represented in dotted, solid and dashed line respectively. Triangles correspond to measurements on STM images.}
\end{figure*}

STM images show that $G\_3\times3$ islands present superlattices (SLs) of various periodicities in the nanometer range (see Fig. \ref{fig1} (a)). Contrary to graphene on SiC$(0001)$ (Si face), graphene on SiC$(000\overline{1})$ (C face) exhibits a significant rotational disorder, already from the first graphene layer. This results in  a ring-shaped graphitic signal on LEED patterns (Fig. \ref{fig1} (b)). A previous STM study established that the SL period depends on the orientation angle of the graphene layer with respect to the substrate surface lattice \cite{hiebel:153412}. In this section we present a more quantitative analysis of the SLs that identifies them as moir\'{e} patterns (MPs).

Moir\'{e} patterns arise from a non linear composition of two periodic lattices\cite{Amidror}. They appear as an additional periodic lattice of larger period than the two components. For example, MPs are observed by STM on graphite\cite{moires} and few layer graphene samples with rotational stacking faults\cite{FLG}. They also show up when two lattices of different lattice parameters are superimposed as for graphene monolayer on transition metals\cite{GsurIr, parga:056807, marchini:075429}. For two periodic lattices with reciprocal lattice vector $\mathbf{k_1}$ and $\mathbf{k_2}$ respectively, the resulting MP is characterized by the reciprocal lattice vector\cite{Amidror,GsurIr}:
\begin{equation}
\mathbf{k_M}= \mathbf{k_2} - \mathbf{k_1}.
  \label{equa_moire}
\end{equation}
In the system we consider, the SiC$(3\times3)$ lattice parameter being almost 4 times bigger than the one of graphene, high order spectral components have to be considered. As we can see on the LEED pattern in Fig. \ref{fig1} (b), first order SiC$(3\times3)$ spots are located far away from the graphitic signal. According to equation (1), moir\'{e} patterns constructed on these spots and any graphene spot would have a smaller period than the SiC$(3\times3)$ lattice, which cannot explain the observed SLs. SiC$(1\times1)$ spots can also be ruled out for similar reasons. From the LEED pattern of fig. \ref{fig1} b), the reciprocal lattice vectors of the SiC$(3\times3)$ reconstruction most likely to lead to MPs with periods in the nanometer range are the high order $(4,0)$, $(3,1)$, $(2,2)$ ones and their symmetric counterparts in the reciprocal space. Fig. \ref{fig1} (c) provides an illustration of a MP construction associated to the $(2,2)$ SiC$(3\times3)$ spot of the LEED pattern, for a graphene island of orientation $\alpha$ with respect to the SiC surface lattice.

Thus, we calculated the moir\'{e} periodicity $P$ as a function of the graphene orientation angle $\alpha$ with respect to the SiC surface lattice with $\alpha$ ranging from $0^\circ$ to $30^\circ$ (due to the symmetry of the system seen on the LEED pattern). For each of the three relevant SiC$(3\times3)$ Fourier components, we use equation (\ref{equa_moire}) and $P(\alpha) =2 \pi / (k_M (\alpha) \cos(\pi / 6))$. The three resulting $P(\alpha)$ curves are plotted in Fig. \ref{fig1} (d). We have also  measured moir\'{e} periodicities versus graphene orientation angles on STM images of monolayer $G\_3\times3$ islands (such as Fig. \ref{fig1} (a)), with an accuracy of $\pm \ 0.1$ nm and $\pm\ 1\ ^\circ$ respectively. As represented on Fig. \ref{fig1} (d) , experimental data do fit very well with calculations. For a given angle, the largest period - which corresponds to the best match in reciprocal space - is generally predominant in the images. We also note that most studied islands exhibit an orientation angle between $15^\circ$ and $30^\circ$. This is consistent with the peculiar rotation angle distribution revealed by LEED\cite{hiebel:153412, revstarke}. We will thus concentrate on these orientations in the following.
To summarize, we interpret superlattices on $G\_3\times3$ monolayer islands as high order MPs, resulting from the superposition of the SiC$(3\times3)$ and the graphene-like lattices. Note however that the moir\'{e} interpretation is essentially geometric and that it does not give any information on the nature of the interaction between graphene and its substrate. We shall consider this point in section \ref{interaction}.

\begin{figure*}[h!]
\includegraphics[width=0.5\textwidth]{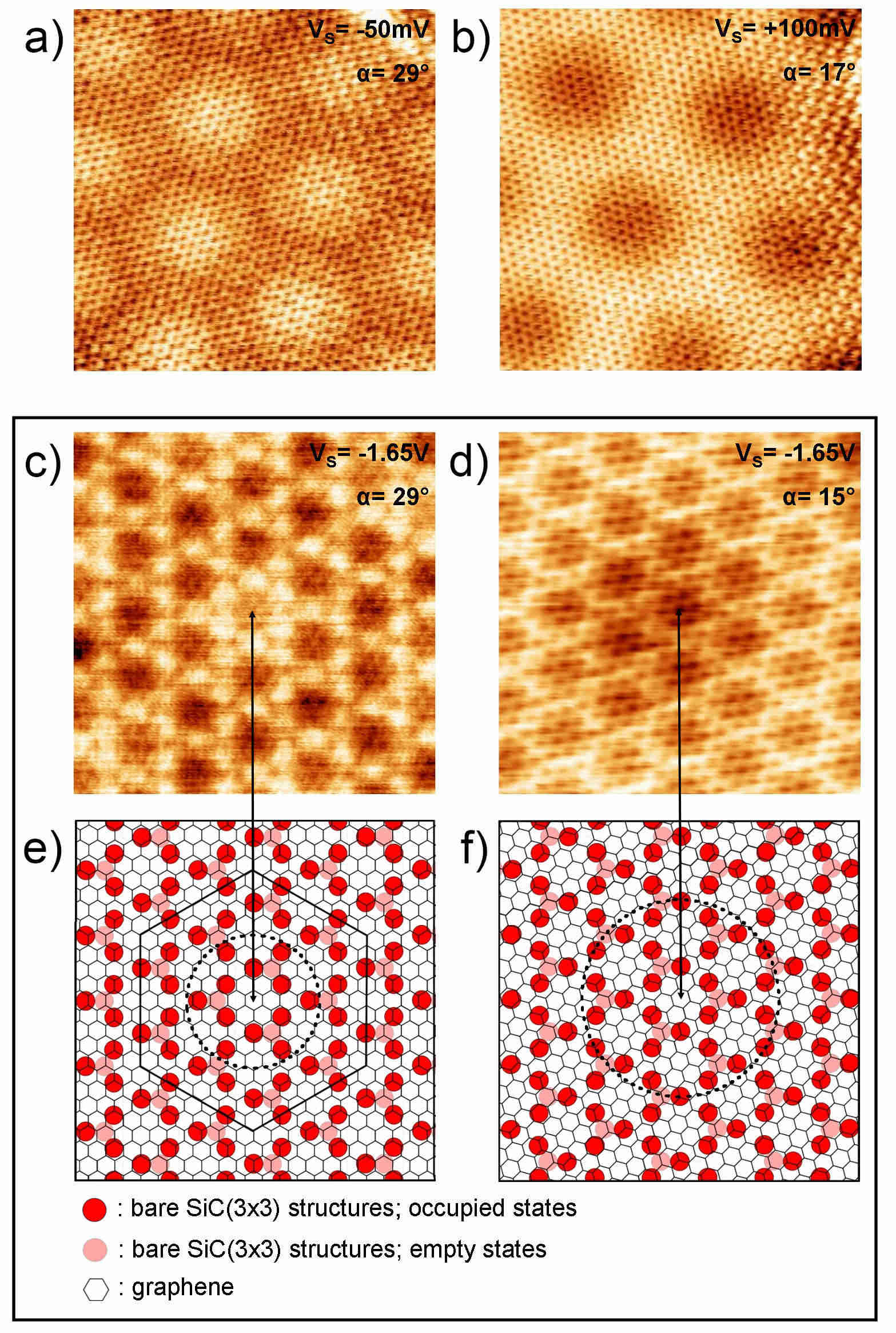}
\caption{\label{fig2}(Color online) (a), (b): $8\times8$ nm$^2$ STM images of $G\_3\times3$ islands with atomic resolution on the graphene layer. Two different moir\'{e} patterns are observed: (a) shows a superstructure with ``ball-like'' contrast (P= $2.9$ nm), typical of a moir\'{e} pattern constructed on the $(2,2)$ SiC$(3\times3)$ Fourier component. (b) shows a superstructure with ``hole-like'' contrast (P= $3.5$ nm), typical of a moir\'{e} pattern constructed on the $(3,1)$ SiC$(3\times3)$ Fourier component. (c), (d): $5\times5 $ nm$^2$ STM images of two other $G\_3\times3$ islands showing the same type of moir\'{e} pattern as (a) and (b) resp. ((c) P= $2.9$ nm (d) P= $4.7$ nm). The sample bias is chosen so that both graphene and SiC$(3\times3)$ lattices are detected and their stacking is thus visible.  (e), (f): Schematic representations of the stacking of the graphene atoms on the substrate reconstruction deduced from images (c) and (d) respectively (the link between the STM images and the schematics is indicated by arrows). On (e), the Wigner-Seitz (pseudo) common cell is represented in solid line. On both illustrations, the dotted circle separates regions of the moir\'{e} cell with different types of stacking.
}
\end{figure*}

We now focus on the atomic structure and on the stacking for graphene islands with a MP constructed on the $(2,2)$ and $(3,1)$ SiC$(3\times3)$ Fourier components which are the most common on our samples ($15^\circ<\alpha<30^\circ$). As shown on Fig. \ref{fig1} (d), the corresponding  moir\'{e} periodicity is maximum for a graphene orientation angle $\alpha$ of $30^\circ$ and $13,9^\circ$ respectively. Low bias STM images (see Fig. \ref{fig2} (a), (b)) show that the MPs observed for angles close to these two values exhibit inverted contrasts: ``ball-like'' for $\alpha$ close to $30^\circ$, ``hole-like'' for $\alpha$ close to $14^\circ$. At the atomic scale, a well defined honeycomb pattern characteristic of monolayer graphene is observed at low bias for both orientations (see Fig. \ref{fig2} (a), (b)). We point out that the moir\'{e} contrast on $G\_3\times3$ islands shows no variations with the tip and tunneling conditions: bright areas remain bright at any bias and for all tips tested (see Fig. \ref{fig5}).

In order to understand the variation of the MP contrast with angle $\alpha$, we studied the local stacking of the graphene and SiC$(3\times3)$ lattices for $\alpha$ close to $30^\circ$ and $14^\circ$. As already mentioned in previous papers\cite{riedl:245406, mallet:041403, hiebel:153412, rutter:235416}, graphene appears transparent on high bias STM images so that the interface -the SiC$(3\times3)$ reconstruction in the present case- becomes visible. Conversely, atomic resolution on graphene is obtained on low bias images. Thus, stacking can be observed using two different approaches: by dual bias imaging at low and high bias or by imaging at an intermediate tunnel bias voltage, which corresponds to a crossover between these two extreme situations (to be discussed in section \ref{electronic}). The latter type of image is represented in Fig. \ref{fig2}(c) and \ref{fig2}(d) for islands with $\alpha= 29^\circ$ and $15^\circ$ respectively. For this sample bias ($V_S=-1.65$ V), the graphene and SiC$(3\times3)$ lattices appear simultaneously. Schematic reproductions of the images in Fig. \ref{fig2}(e) and (f) give a clear view of the local stacking (the same result is found from dual bias imaging). Since no established structure model for the SiC$(3\times3)$ reconstruction exist, we represent in Fig. \ref{fig2} (e), (f) the states detected by STM on the substrate surface\cite{Hoster3x3}: filled (empty) states are represented in red (light-red) (dark gray and light gray in the printed version).

In the $\alpha = 30^\circ$ case (Fig \ref{fig2} (e)), the graphene and SiC$(3\times3)$ lattices are quasi commensurate. The corresponding common Wigner-Seitz cell is represented by solid lines. The apparent height of the MP is maximum in the center of the cell and minimum on its edges. These areas correspond to two different types of stacking. At the center of the cell (circled area), SiC$(3\times3)$ states  are located under the center of graphene hexagons (i.e. no C atom is in coincidence with SiC$(3\times3)$ states). On the edges (dark area), all SiC$(3\times3)$ states have C atoms or C-C bonds on top of them.

In the $\alpha = 15^\circ$ case, the moir\'{e} corrugation is inverted. The apparent height of the moir\'{e} is minimal at the center of the cell and maximal at its edges. Now the stacking at the center of the cell (circled area) is such that every  SiC$(3\times3)$ state has C atoms or C-C bonds directly above which is similar to the stacking in the dark area of the $\alpha \approx30^\circ$ case. At the edges of the cell, a significant amount of SiC$(3\times3)$ states are located under the center of graphene hexagons, as for the bright regions in the $\alpha \approx30^\circ$ case. Therefore, the local stacking of bright (high) and dark (low) areas is the same for the two kinds of MP contrast, ``ball-like'' ($\alpha\ \approx 30^\circ$) and `` hole-like'' ($\alpha\approx 14^\circ$). The apparent MP ``contrast inversion'' arises from changes in the local stacking induced by the graphene rotation.

Islands with $\alpha \approx\ 30^\circ$ deserve particular attention because they allow a direct comparison with the experimental results for the Si face and with theoretical calculations. For $\alpha=30^\circ$, the graphene and the SiC$(1\times1)$ lattices are (quasi)-commensurate with a  $6\sqrt{3}\times6\sqrt{3}R(30^\circ)$-SiC ($6R3$) common cell (or a $(13\times13)$ graphene cell). This is the configuration which is observed for the Si face\cite{vanBommel, PhysRevB.58.16396,Tsai},  
the layer orientation is imposed by the substrate and is therefore the same on the whole sample. A strong interaction between the first graphitic layer (``buffer layer'') and the substrate occurs so that only the second layer shows graphene properties \cite{varchon:126805, kimPRL, varchon:235412, mattausch:076802}. In particular, no honeycomb contrast characteristic of graphene has ever been observed in STM studies of the $6R3$ phase of the Si face -corresponding to the first graphitic layer or ``buffer layer''- since it lacks $\pi$  states in the vicinity of the Fermi level\cite{emtsev:155303}. Additionally, this $6R3$ usually gives rise to a dominant  SiC$(6\times6)$ superstructure in STM images\cite{Tsai}, although high resolution images reveal the actual $6R3$ periodicity\cite{riedl:245406}. 

\begin{figure*}[h!]
\includegraphics[width=0.5\textwidth]{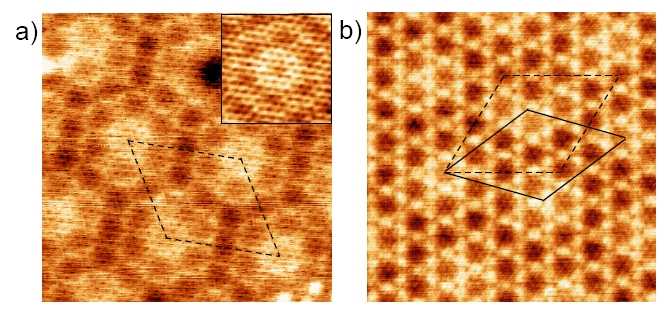}
\caption{\label{fig3}(Color online) $8\times8$ nm$^2$ STM images of (a) an $\alpha= 30^\circ$ $G\_3\times3$ island; sample bias: $-1.5$ V. The SiC$(6\sqrt{3}\times6\sqrt{3})R30^\circ$ common cell is represented with dashed lines. Insert: $3\times3$nm$^2$ STM image of the same island; sample bias: $+100$ mV.The graphene layer shows an AB symmetric honeycomb contrast. (b) an $\alpha= 29^\circ$ $G\_3\times3$ island; sample bias: $-1.65$ V. The SiC$(6\sqrt{3}\times6\sqrt{3})R30^\circ$ cell represented by dashed lines clearly does not coincide with the moir\'{e} pseudo unit cell represented by solid lines.
}
\end{figure*}

A totally different situation occurs for graphene on the SiC$(3\times3)$ reconstruction of the C face. For $\alpha=30^\circ$, Fig. \ref{fig3} (a), we actually observe a $2\sqrt{3}\times2\sqrt{3}R(30^\circ)$ with respect to the SiC$(3\times3)$, which corresponds to the actual $6R3$ (and not SiC$(6\times6)$) superstructure with respect to the SiC $(1\times1)$. More important, the honeycomb contrast of graphene clearly shows up at low bias (see inset in Fig. \ref{fig3}). This demonstrates that the graphene states are present close to the Fermi level and thus implies a comparatively much weaker interaction with the substrate compared to the Si face. This weak coupling probably results from the presence of the SiC$(3\times3)$ surface reconstruction below the graphene layer. Indeed, ab-initio calculations performed for a graphitic C layer on the ideal (non-reconstructed) C face for this orientation ($\alpha=30^\circ$) indicate a strong bonding to the substrate\cite{varchon:126805, varchon:235412, mattausch:076802} and subsequently the disappearance of the $\pi$ states at low energy, as for the Si face. This suggests that the SiC$(3\times3)$ reconstruction efficiently passivates the substrate surface for the C face, preventing the formation of chemical bonds with the graphene layer. Similar results were obtained from ab-initio calculations for graphene monolayer on the SiC$(2\times2)_C$ reconstruction\cite{magaud:161405}, a system that coexists with $G\_ 3\times3$ on the SiC$(000\overline{1})$ graphitized surface.

Another observation we have made on $G\_ 3\times3$ islands for $\alpha \approx\ 30^\circ$ is that the orientation of the graphene layer is not locked to $30^\circ$. To see that, we took advantage of the fact that the MP orientation with respect to the SiC lattice is highly sensitive to the rotation angle $\alpha$. Indeed, for a change by $1^\circ$ of $\alpha$, the orientation of the MP changes by $13^\circ$ (for $\alpha \approx\ 30^\circ$). Thus, we could identify some slight deviations ($\Delta \alpha < 1^\circ$) from the quasi-commensurate ($\alpha=30^\circ$) configuration. Fig. \ref{fig3} (b) provides an illustration of this effect: the MP significantly differs from the $6R3$ supercell although $\alpha$ remains close to $30^\circ$ (we measure $\alpha=29\pm 1^\circ$). The fact that such islands exist suggests that the $6R3$ (quasi-commensurate) configuration does not lead to a notable energy reduction, at variance with the case of the Si face. This is consistent with the absence of covalent, directional bonds between graphene and substrate at the interface for $G\_ 3\times3$ islands.

\subsection{\label{electronic}Electronic structure of monolayer graphene on the SiC$(3\times3)$ reconstruction}

\begin{figure*}[h]
\includegraphics[width=0.5\textwidth]{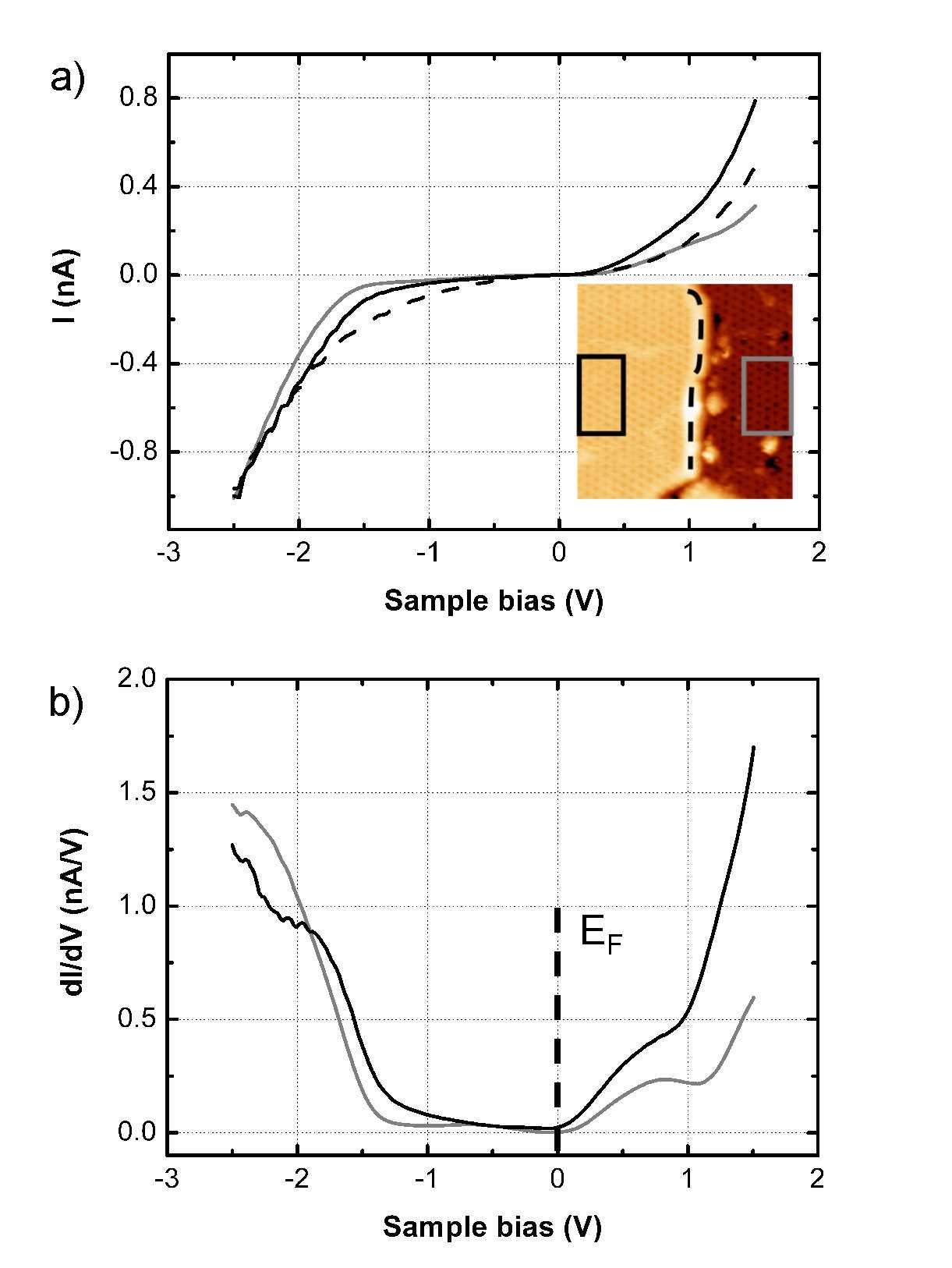}
\caption{\label{fig4} (Color online) (a) Inset: $20\times20$ nm$^2$ STM image of a region with the bare SiC$(3\times3)$ reconstruction (right side) and a $G\_3\times3$ island (left side). Sample bias: $-2.5$ V. The I(V) curves are spatially averaged over the boxed regions ($300$ points) on the bare SiC$(3\times3)$ reconstruction (gray line) and on the $G\_3\times3$ island (black line) and over $15$ points of the island edge (dashed line). For STS measurements: Setpoint voltage: $-2.5$ V; Setpoint current: $1.0$ nA.
(b) Corresponding dI/dV curves for the $G\_3\times3$ island (black line) and the bare SiC$(3\times3)$ reconstruction (gray line).}
\end{figure*}

After these mainly structural considerations, we focus on the electronic structure of the graphene overlayer and of the SiC$(3\times3)$ reconstruction. We first compare the electronic structure of the $G\_3\times3$ and SiC$(3\times3)$ phases using current imaging tunneling spectroscopy (CITS). This technique consists in acquiring a constant current image and an I(V) curve after each of its points. For each spectrum, the feedback loop is turned off and the sample voltage ($V_S$) is ramped between preset values. I(V) curves are then numerically differentiated to get dI/dV conductance curves which are -in first approximation- proportional to the local density of states (LDOS) of the sample surface.
In Fig. \ref{fig4}, we present CITS data acquired on a region (see insert in Fig. \ref{fig4} (a)) with the bare SiC$(3\times3)$ reconstructed substrate surface (right) and a $G\_3\times3$ island (left), so that both region are probed with the same tip. Fig. \ref{fig4} (a) shows three I(V) curves, one for each type of surface, spatially averaged over the boxed regions (300 points each), and one for the edge of the graphene island (averaged over 15 points). The dI/dV spectra for the $G\_3\times3$ island and the bare substrate are given in Fig. \ref{fig4} (b). 

For the bare SiC$(3\times3)$ reconstruction, the I(V) curve in Fig. \ref{fig4} (a) displays a dramatic reduction of the current between $V_S=-1.4$V and $+0.1$V. This feature is still visible -although less marked- in the I(V) curve for the graphene island. The curve obtains on the edge of the island indicates that the lack of current at low bias does not arise from the electronic structure of the tip. These observations suggest the presence of a surface bandgap associated to the SiC$(3\times3)$ reconstruction that subsists under the graphene layer. However, a residual current related to in-gap states is detected in the surface gap of the bare SiC$(3\times3)$ reconstruction (between $V_S=-1.4$V and $-0.5$V). To further study the electronic structure of the $G\_3\times3$ island and of the bare substrate, the conductance curves presented in Fig. \ref{fig4} (b) are analysed in the following. 

The SiC$(3\times3)$ spectrum exhibits a region of minimum conductance ranging from $V_S=-1.4$ V to $+0.1$ V. These values are only weakly dependent (within $0.2$ V) of the tip and sample. Hence, the SiC$(3\times3)$ reconstruction presents an asymmetric surface bandgap, with the Fermi level close to the bottom of the conduction band, as expected for a n-type semiconductor.
A broad structure centered around $V_S=-0.7$ V is also detected. It is ascribed to in-gap states. An additional CITS study of the SiC$(3\times3)$ reconstruction suggests that they arise from a subsurface atomic layer (not shown). For the occupied states, these observations are consistent with Angle Resolved Photoemission Spectroscopy (ARPES) data\cite{emtsev:155303} of the bare and lightly graphitized SiC$(000\overline{1}) (3\times3)$ surface where a large intensity for binding energies larger than $1.5$ eV and a residual emission between $0.5$ eV and $1.0$ eV binding energies are detected.

The $G\_3\times3$ spectrum has similarities with the SiC$(3\times3)$ spectrum. In particular, the structure at $V_S=+0.7$ V (above the top of the substrate surface gap) and the rapid increase of conductance below the bottom of the substrate surface gap ($V_S\approx -1.4$ V) are clearly observed. These structures arise from the underlying SiC$(3\times3)$ reconstruction. The Fermi level of graphene ($E_F$ in Fig.\ref{fig4} (c)) is located close to the top of the substrate surface gap. Inside the SiC$(3\times3)$ surface bandgap, an additional - though rather small - density of states originating from the graphene layer is detected. In other words, outside the (bare) SiC$(3\times3)$ surface bandgap, the signal is dominated by the contribution of the substrate which explains the transparency of graphene at high bias\cite{riedl:245406, mallet:041403, hiebel:153412, rutter:235416}.
 
Note that the surface bandgap of the substrate remains unchanged below the graphene layer. This again suggests a weak graphene - substrate interaction since strong coupling would also affect the electronic states of the SiC$(3\times3)$ reconstruction. Moreover, for an extended energy range within the surface bandgap (from $-1.4$ eV to $+0.1$ eV) the density of interface states susceptible to interact with graphene states is quite small (from Fig. \ref{fig4}). This is consistent with graphene-like atomic contrast on low bias STM images presented here (Fig. \ref{fig2}) and in previous papers \cite{hiebel:153412}. 
Nevertheless, moir\'{e} patterns are still visible on the graphene islands at energies within the SiC$(3\times3)$ surface bandgap. This gives evidence for a residual effect of the substrate. Following experiments aim at discriminating between a topographic or an electronic effect for the MP contrast.

\begin{figure*}[h]
\includegraphics[width=0.5\textwidth]{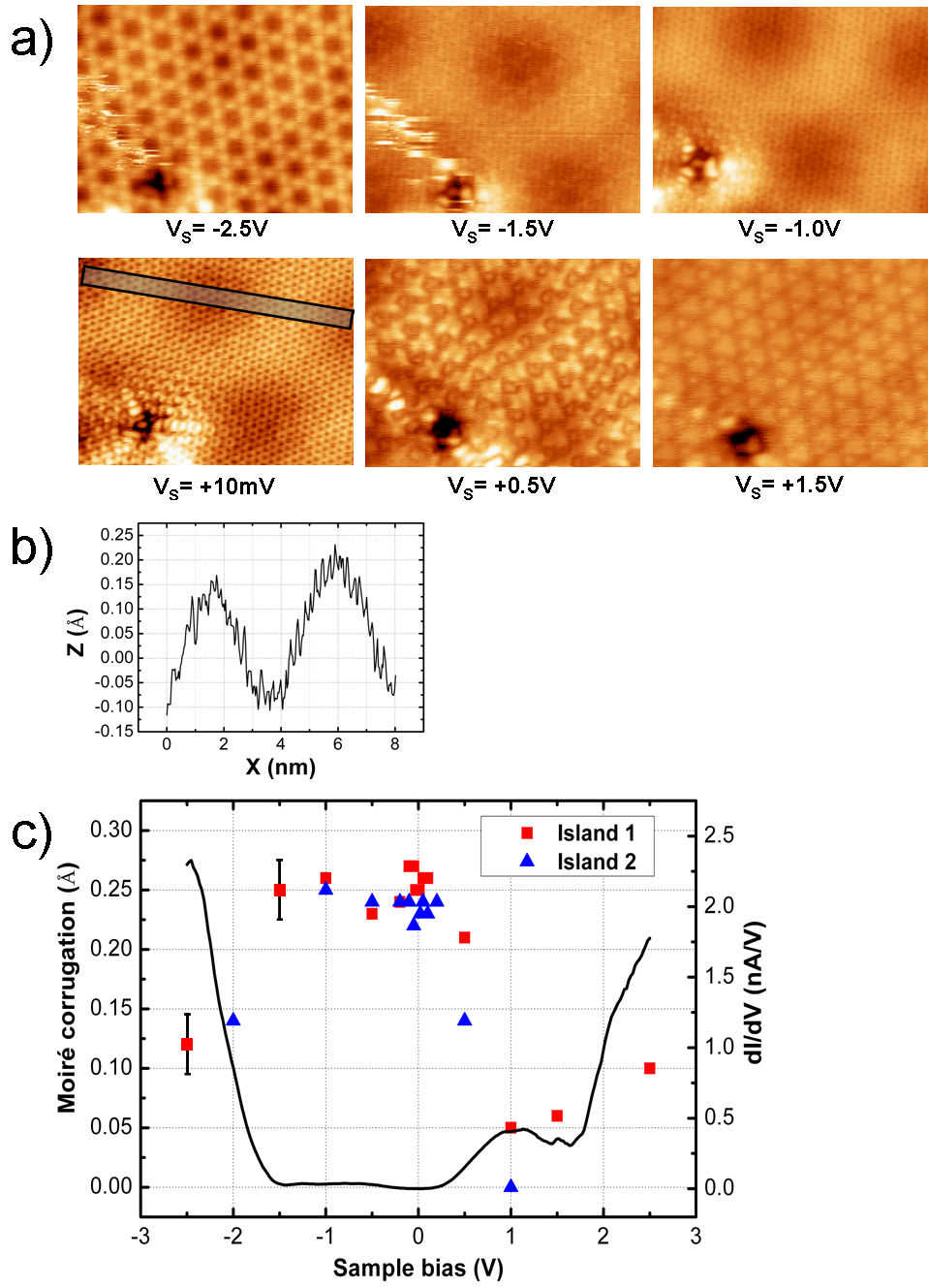}
\caption{\label{fig5} (Color online) (a) A series of $8\times6$ nm$^2$ STM images taken at the same spot of a $G\_3\times3$ island ($\alpha = 16 ^\circ$, $ P= 4.5$ nm) at various sample biases ($V_S$)(Common height range $\Delta Z=1$\AA). The relative amplitude of the signals arising from the SiC$(3\times3)$ reconstruction and from the graphene layer changes with sample bias. (b) Laterally averaged profile taken over the boxed region in (a) for $V_S=+10$mV. (c) The peak-to-peak moir\'{e} corrugation amplitude as a function of the sample bias voltage, on two similar $G\_3\times3$ islands ($\alpha = 16 ^\circ$, $ P= 4.5$nm)  of two different samples obtained with different tips. The island referred to as ``island 1'' is the island presented in (a). A dI/dV curve acquired on the bare SiC$(3\times3)$ reconstruction is also represented on the graph (solid line) in order to locate the substrate surface bandgap. The moir\'{e} corrugation and the LDOS of the substrate show complementary behaviors.
}
\end{figure*}

 In Fig. \ref{fig5} (a) we present a series of STM images of the same area of a $G\_3\times3$ island with $\alpha = 16 ^\circ $\ at various sample bias voltages. At $V_S=-2.5$ V, the SiC$(3\times3)$ reconstruction is clearly visible while no evident graphene signal is detected. At $V_S=-1.5$ V, atomic resolution on graphene appears, superimposed to the SiC$(3\times3)$ signal as in Fig. \ref{fig2}(c), \ref{fig2}(d) and \ref{fig3}. For lower biases, typically from $V_S=-1.0$ V to $+0.2$ V, we detect a well defined honeycomb graphene signal and no more signal of the underlying reconstruction\cite{contrast} (see $V_S=-1.0$ V and  $V_S=+10$ mV panels). From $V_S=+0.5$ V to higher biases, no more evident graphene signal is visible and the SiC$(3\times3)$ signal reappears. Thus, atomic resolution on graphene is mostly achieved within the SiC$(3\times3)$ surface bandgap, as expected since the DOS arising from the SiC$(3\times3)$ is small within the surface bandgap. If we now concentrate on the moir\'{e} pattern - the signal with period $P=4.5$ nm in the images of Fig. \ref{fig5} (a) - we notice that its corrugation is obviously much smaller at high biases than	at low biases. 

To complete these observations, we have measured the peak-to-peak moir\'{e} corrugation amplitude as a function of the sample bias voltage (see graph in Fig. \ref{fig5} (c)) using two different methods: i) profiles laterally averaged over 0.6nm on raw images or ii) profiles taken on low-pass filtered images in order to get rid of the high frequency atomic corrugation. Both methods gave the same results. The former method is illustrated in Fig. \ref{fig5} (b) for a profile taken on the low bias image ($V_S=+10$ mV) of Fig. \ref{fig5} (a). We stress that the measurements reported in Fig. \ref{fig5} (c) were actually made at several spots on larger images. The uncertainty in the measurement is estimated to be of $\pm 0.025$ \AA \ . It results from the residual contribution of the atomic corrugation (SiC$(3\times3)$ at high bias or graphene at low bias as in Fig. \ref{fig5} (b)) and to some inhomogeneities in the corrugation of the moir\'{e} patterns. The data presented in Fig. \ref{fig5} (b) were acquired with different tips on two different $G\_3\times3$ islands (labeled ``Island 1'' and ``Island 2'') of approximately the same orientation: $\alpha =16 ^\circ$, $P=4.5$ nm (STM images in Fig. \ref{fig5} (a) were obtained  on ``Island 1''). A typical SiC$(3\times3)$ average spectrum is also given on the graph in order to locate the substrate surface bandgap. The behavior is similar for both set of measurements with the following characteristics: within the SiC$(3\times3)$ surface bandgap, the moir\'{e} corrugation amplitude is constant and equal to $0.25 \pm 0.025 $ \AA . Outside the surface bandgap, when the SiC$(3\times3)$ contribution to the tunneling current becomes dominant, the corrugation dramatically decreases -or even vanishes. Other $G\_3\times3$ islands with different orientations (including angles close to $30^\circ$) show similar behavior of the  moir\'{e} corrugation as a function of bias. These observations imply that the moir\'{e} corrugation is associated to the graphene layer and not to the SiC$(3\times3)$ reconstruction\cite{moire}. 
Moreover, since the corrugation is independent of the bias in a large voltage range spanning the SiC$(3\times3)$ surface bandgap, it is most probably of topographic origin. The corrugation amplitude varies with the moir\'{e} period, between $0.15$ \AA \ and $0.25$ \AA \ (smaller corrugations correspond to moir\'{e} patterns of smaller periods).

\begin{figure*}[h]
\includegraphics[width=0.5\textwidth]{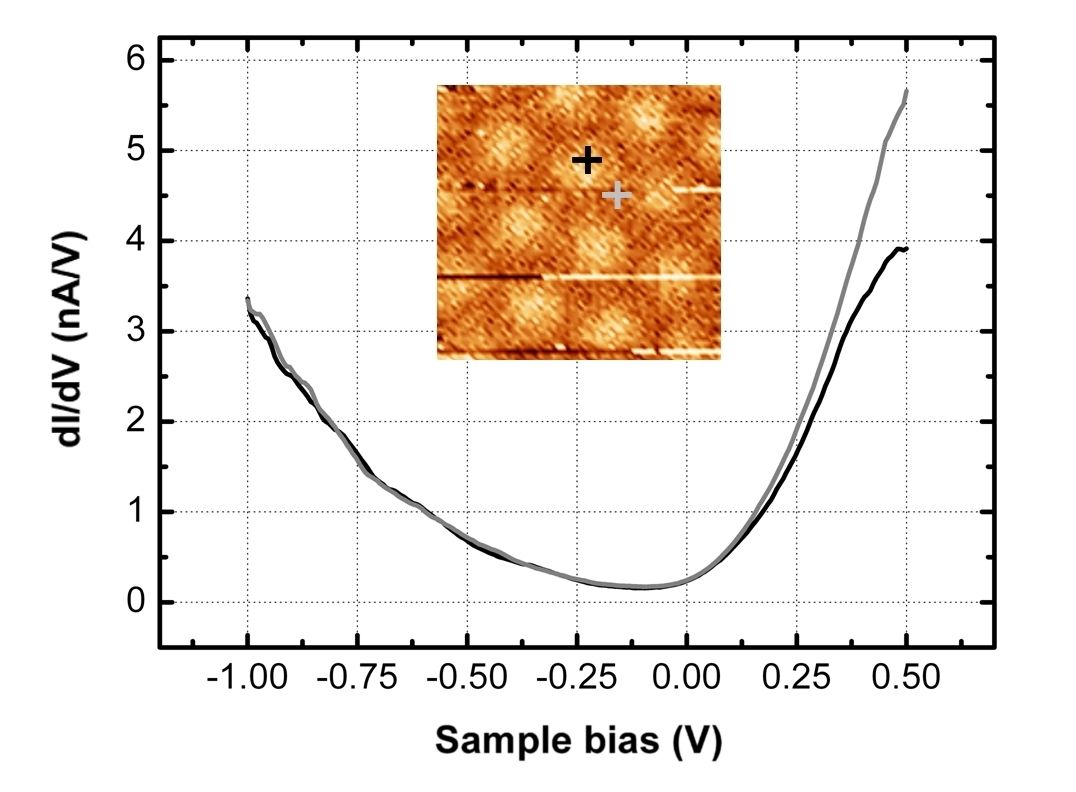}
 \caption{\label{fig6} (Color online) Inset: $10\times10$ nm$^2$ STM image of a $G\_3\times3$ island presenting a ``ball-like'' moir\'{e} contrast. ($\alpha = 28 ^\circ$, $ P= 2.9$ nm). Sample bias: $-1.0$ V. dI/dV curves are spatially averaged on $24$ points of highest topographic regions (black) and of lowest topographic regions (gray). Setpoint voltage: $-1.0$ V; Setpoint current: $1.0$ nA. No spatial variation in the electronic structure of graphene is observed in the voltage range $[-1.0$ V$, +0.15$ V$]$.}
\end{figure*}

From the above observations, the graphene-substrate distance changes by $0.2$ \AA \ from the highest to the lowest areas of the moir\'{e} pattern. To look for a possible change in the electronic properties correlated to these soft ``ripples'', we have performed scanning tunneling spectroscopy on several islands. In Fig. \ref{fig6} we present CITS results on a $G\_3\times3$ island with $\alpha\approx 30^\circ$ (see insert) - similar results were obtained for other values of $\alpha$ between $15^\circ$ an $30^\circ$. The setpoint was chosen within the substrate surface bandgap ($V_S=-1.0$ V) with a setpoint current of $1.0$ nA in order to probe essentially the graphene states. The graph shows two spectra, one corresponding to the highest regions of the moir\'{e} pattern and the other to lowest regions. Both are averaged over 24 points. The spectra coincide from $V_S=-1.0$ V to $+0.15$ V. This demonstrates that the electronic structure of the graphene overlayer is homogeneous and thus not affected by the moir\'{e} pattern in a wide energy range spanning the Fermi level (and located in the substrate surface bandgap). Note however that the spectroscopy measurements are conducted at room temperature and the energy resolution is thus limited to $\approx 0.1$ eV\cite{0295-5075-61-3-375}. From $V_S=+0.2$ V to $+0.5$ V, more signal is detected on the low regions than on the high ones. This discrepancy arises from topographic effects (decrease in MP corrugation) discussed in connection with Fig. \ref{fig5}. 

Another important question is the position of the Dirac point. Previous ARPES\cite{emtsev:155303} and transport measurements\cite{CB} assess that it is located 0.2 eV below the Fermi level. But these techniques are non-local and give therefore an average value of the doping of the graphene layer. Importantly, the underlying reconstruction of the substrate is not identified in the probed region. STS is thus in principle the most adapted technique for answering this question. In Fig. \ref{fig6}, we find a rather structureless spectrum with a ``flat'' minimum ranging from $V_S=-0.2$ V to $0$ V. Some other spectra showed a well defined minimum located around $V_S=-0.25$ V. However, due to a significant variability in our measurements of the dI/dV curves between $V_S=-0.5$ V and $0$ V, we refrain from giving a definite value for the position of the Dirac point.

\subsection{\label{interaction}Graphene on $6H$-SiC$(000\overline{1})(3\times3)$: an almost ideal graphene layer?}

From measurements presented in Fig. \ref{fig5}, we find a topographic corrugation of $0.15$ \AA \ to $0.25$ \AA \  of the graphene monolayer while no long range topographic modulations were observed on the bare SiC$(3\times3)$ reconstruction. The period of the graphene topographic modulation is related to its orientation with respect to the substrate reconstruction and follows a moir\'{e} model (discussed in connection with Fig.\ref{fig1}). This means that the graphene corrugation is induced by the SiC$(3\times3)$ reconstruction. More precisely, the graphene - substrate distance is governed by the local stacking of the SiC$(3\times3)$ and the graphene overlayer as shown in section \ref{geom}. 

Such an effect has already been reported for graphene on transition metals. It is instructive to compare our data with a well documented case of a relatively strong coupling, such as graphene on Ru$(0001)$, where the $\pi$ bands of graphene are strongly perturbed by interaction with the substrate\cite{brugger:045407, SutterNano}. This system presents a moir\'{e} pattern ($P=2.9$ nm) caused by the lattice mismatch between graphene and Ru. A signal with the periodicity of graphene is observed by STM \cite{parga:056807, marchini:075429} but the contrast changes from honeycomb in the high region to triangular in the low areas\cite{parga:056807, marchini:075429, sutter:133101}. This is at variance with the uniform honeycomb pattern we observe on $G\_3\times3$. DFT calculations \cite{wintterlinPCCP}, surface X-ray diffraction\cite{martoccia:126102}, STM\cite{parga:056807, sutter:133101} and core level spectroscopy \cite{preobrajenski:073401} conclude that lower areas of the graphene layer strongly bond to the substrate. 

STS results presented for graphene on Ru$(0001)$ in Ref. \onlinecite{parga:056807} are of particular relevance for the purpose of our study. It shows dI/dV spectra with significant spatial variations correlated to the moir\'{e} pattern. This finding was interpreted in Ref. \onlinecite{parga:056807} using a generic model where a periodic potential -with the periodicity of the MP- is applied to a flat and isolated graphene layer. This leads to the LDOS modulations observed by STM, which also corresponds to charge inhomogeneities in the graphene layer. In Ref. \onlinecite{wang:099703}, it was inferred from DFT calculations that the spatial variations of the STS spectra should be attributed to the spatially heterogeneous bonding between graphene and Ru. This is clearly different from the behavior we observe for the dI/dV spectra on $G\_3\times3$ (Fig. \ref{fig6}), and we thus conclude that neither charge modulation nor local (periodic) bonds formation occurs in this system, whatever the orientation angle $\alpha$. Incidentally, even for graphene on SiC$(0001)$, where the graphene overlayer (second graphitic plane) is known to be well decoupled from the substrate, spatial variations of the dI/dV spectra have been reported close to the Dirac point \cite{vitali}. Therefore graphene on the $G\_3\times3$ islands may be quite close to ideal graphene, due to a weak interaction with the substrate reconstruction.

We now briefly discuss the origin and the influence of the corrugation of the graphene layer which give rise to the MP. Since strong periodic bonding to the substrate can be ruled out from our data, these topographic modulations probably come from a weak, possibly Van der Waals-like, interaction that depends on the local stacking. Note that the corrugation we measure is small, typically $0.2$ \AA \ Peak to Peak (PP) for wavelengths $P$ in the range $2-5$ nm.
The consequence of such ``ripples'' on the electronic structure of isolated graphene layers has been estimated in previous papers. 
For a graphene layer with a modulation of pseudo-period $P'=1.9$ nm and an amplitude of 0.4 \AA\ PP, DFT calculations \cite{kimPRL, varchon:235412} show no significant modification of the electronic properties with respect to the flat configuration.
In particular, it does not open a gap at the Dirac point\cite{kimPRL}. 
Even on an isolated strongly corrugated monolayer ($1.5$ \AA\ PP for a period $ P \approx 3 $ nm), other ab-initio calculations shows that the LDOS of graphene remains linear within $\pm 1$ eV from the Dirac point in the high (and low) regions \cite{wang:099703}. 
Thus we believe that the small topographic corrugation we observe should have only a limited effect on the electronic structure of graphene close to the Dirac point for $G\_3\times3$ islands. 
However, experiments with an improved resolution should be performed to search for -or to rule out- a possible influence of the superperiod (MP) on the band structure of graphene\cite{louieNatMat, pleti}. 

\section{Conclusion}
We have investigated by STM and STS graphene monolayer islands grown under UHV on the SiC$(3\times3)$ reconstruction of the $6H$-SiC$(000\overline{1})$ surface. These islands present different orientations with respect to the substrate. From STM topographic images with atomic resolution, we find that the various superstructures with periods in the nm range observed on the islands can be interpreted as moir\'{e} patterns arising from the composition of graphene and high order SiC$(3\times3)$ lattice Fourier components. We show that the moir\'{e} contrast corresponds to topographic modulations in the graphene layer of typically $0.2$ \AA. The local graphene-substrate stacking in the low and high regions of the moir\'{e} pattern could be obtained using the transparency of the graphene in high-bias images.
Our STS study of the SiC$(3\times3)$ substrate reconstruction reveals a surface bandgap (typically ranging from -1.4 eV to +0.1 eV) that persists under the graphene monolayer. This characteristic explains the variations with sample bias voltage of the graphene/substrate signal ratio for STM topographic images and for STS. 
Further STS measurements show that the electronic structure is spatially homogeneous for any orientation of the graphene layer which indicates a weak graphene substrate interaction. This is confirmed by the absence of preferential graphene orientations even for an almost commensurate configuration ($\alpha \approx 30^\circ$).
The weak interaction is achieved thanks to the surface reconstruction that efficiently passivates the SiC substrate since ab initio calculations for a bulk-truncated SiC$(000\overline{1})$ surface have predicted strong interaction and covalent bonds formation between the graphene layer and the substrate\cite{varchon:235412, varchon:126805, mattausch:076802}. This suggests that the graphene-substrate coupling can be tuned using post-treatments that alter the substrate surface reconstruction. Finally, in an energy range of $\pm 100$ meV spanning the Fermi energy, very few substrate interface states are susceptible to couple with graphene states which makes the graphene monolayer on SiC$(3\times3)$ a nearly ideal system for investigating low energy excitations.

\begin{acknowledgments}
This work was supported by the French ANR (``GraphSiC'' project N$^\circ$ ANR-07-BLAN-0161) and by the R\'{e}gion Rh\^{o}ne-Alpes (``Cible07'' and ``Cible08''programs). F. Hiebel held a doctoral fellowship from la R\'{e}gion Rh\^{o}ne-Alpes.
\end{acknowledgments}



\begin{thebibliography}{64}
\bibitem[{con()}]{contrast}
\bibinfo{note}{Occasionally, a small contribution from the previously quoted
  in-gap states of the SiC$(3\times3)$ is seen in the range $[-1.0V, -0.5V]$.}
\bibitem[{moi()}]{moire}
\bibinfo{note}{We note that the measured decrease rate of the corrugation
  outside the gap depends on the tip state. These variations can be attributed
  to a varying sensitivity of the tips to interface or graphene states. A weak
  graphene signal superimposed on the substrate-related structures is sometimes
  observed up to 0.5 eV beyond the gap.}
\end{thebibliography}

\end{document}